\documentclass{iopconfser}
\usepackage{xspace}
\usepackage{caption}
\usepackage{subcaption}
\usepackage{cleveref}
\usepackage{graphicx}
\usepackage{url}
\begin{document}

\newcommand{\belle}{Belle~II\xspace}
\newcommand{\legendre}{\textit{Baseline Finder}\xspace}
\newcommand{\cat}{\textit{CAT Finder}\xspace}
\newcommand{\bat}{\textit{BAT Finder}\xspace}
\def \opengen {\ensuremath{\alpha^{3D}_{\mathrm{MC}}}\xspace}
\def \pgen {\ensuremath{p_{\mathrm{MC}}}\xspace}
\def \ptgen {\ensuremath{p_{\mathrm{t,MC}}}\xspace}
\def \thetagen {\ensuremath{\theta_{\mathrm{MC}}}\xspace}
\def \phigen {\ensuremath{\phi_{\mathrm{MC}}}\xspace}
\def \zgen {\ensuremath{z_{\mathrm{MC}}}\xspace}
\def \rhogenthreed {\ensuremath{v_r^{\mathrm{MC}}}\xspace}
\def \rhogen {\ensuremath{v_{\rho}^{\mathrm{MC}}}\xspace}
\def \pyg {\texttt{PyTorch Geometric}\xspace}
\def \pytorch {\texttt{PyTorch}\xspace}
\def \basf {\texttt{basf2}\xspace}
\title{Multi-Modal Track Reconstruction using Graph Neural Networks at Belle II}

\author{Lea~Reuter$^{1}$, Tristan~Brandes$^{1}$, Giacomo~De~Pietro$^{1,2}$ and Torben~Ferber$^{1}$}

\affil{$^1$Institute of Experimental Particle Physics (ETP), Karlsruhe Institute of Technology (KIT),
Karlsruhe, Germany}
\affil{$^2$Scientific Computing Center (SCC), Karlsruhe Institute of Technology (KIT), Karlsruhe,
Germany}

\email{lea.reuter@kit.edu}

\begin{abstract}
High backgrounds and detector ageing impact the track finding in the \belle central drift chamber, reducing both track purity and track efficiency in events. This necessitates the development of new track finding algorithms to mitigate detector performance degradation. 
Building on our previous success with an end-to-end multi-track reconstruction algorithm for the \belle experiment at the SuperKEKB collider (arXiv:2411.13596), we have extended the algorithm to incorporate inputs from both the drift chamber and the silicon vertex tracking detector, creating a multi-modal network. 
We employ graph neural networks to handle the irregular detector structure and object condensation to address the unknown, varying number of particles in each event. 
This approach simultaneously identifies all tracks in an event and determines their respective parameters.
We demonstrate the algorithm's effectiveness using a realistic full detector simulation, which incorporates beam-induced backgrounds and noise modelled from actual collision data. The simultaneous reconstruction of the information from the two detectors yields a track efficiency improvement from 48.0\% to 74.7\% for uniformly displaced particles up to 100\,cm, while increasing the track purity by 5.5~percentage points. We provide a detailed comparison of its track-finding performance against the current Belle II baseline across various event topologies. 
\end{abstract}
\newpage

\section{Introduction}
\label{sec:introduction}
\begin{figure}
    \centering
    \includegraphics[width=0.95\linewidth]{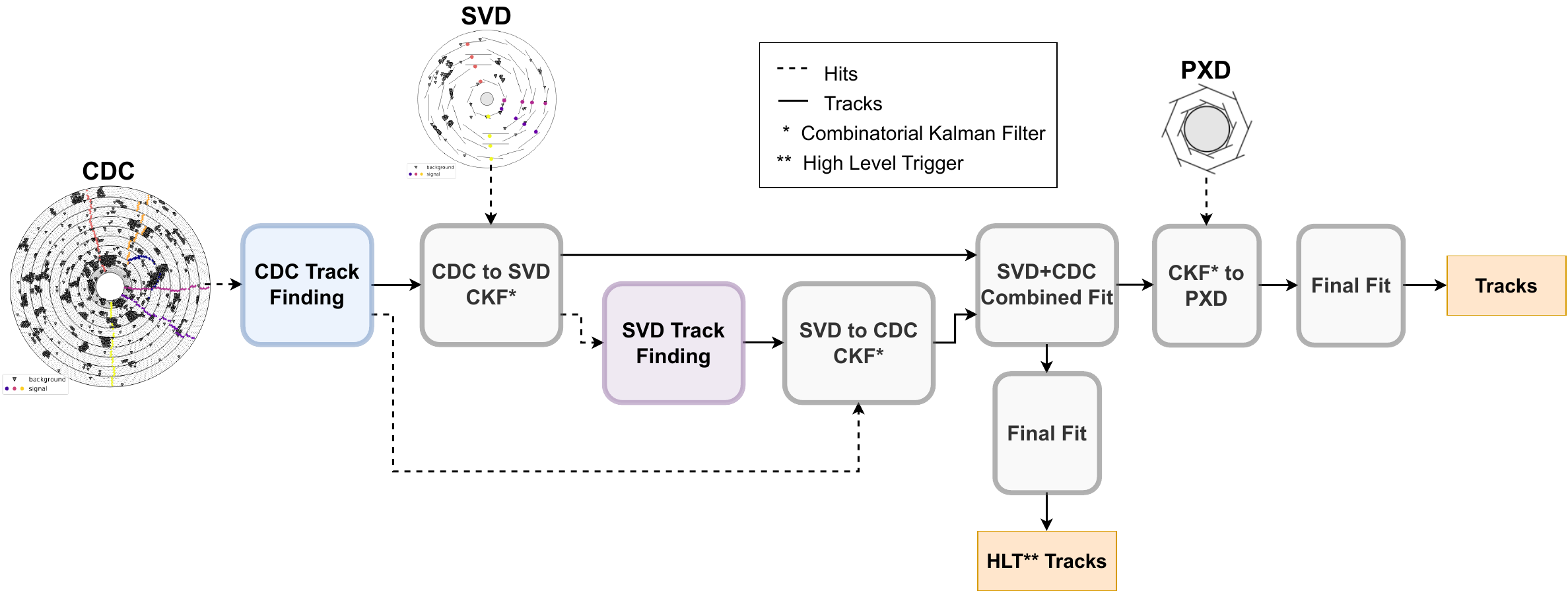}
    \caption{The \belle track finding chain for high level trigger tracks and tracks used in offline reconstruction. The blue box shows the CDC track finding algorithm, while the purple box indicates the SVD track finding algorithm. Final reconstruced objects are highlighted in orange. The inputs of the three tracking detectors CDC, SVD, and PXD are given by one example event display, where the hits from signal particles are shown in colored circular markers and the beam-background is shown with grey triangular markers. See text for details.}
    \label{fig:baseline_tracking}
\end{figure}
The \belle experiment at the SuperKEKB accelerator investigates flavour physics and searches for physics beyond the Standard Model. 
Charged-particle tracking is fundamental for reconstructing particle trajectories and enables precise vertexing, momentum determination, and background suppression. 
The Belle~II tracking system consists, from the beam-pipe outward, of the silicon Pixel Detector (PXD), the Silicon Vertex Detector (SVD), and the Central Drift Chamber (CDC)~\cite{Belle-II:2010dht}.
Due to the PXD’s close proximity to the beamline, the PXD has a very high hit occupancy and is intentionally excluded in the track reconstruction at the High Level Trigger (HLT). 
For this reason, track finding in \belle relies exclusively on hits from the SVD and the CDC.
During the full offline reconstruction, once track candidates have been found and fitted using these detectors, they are extrapolated inward to the PXD, where compatible PXD hits are subsequently associated with the existing track candidates.
Challenges for combined track finding arise because the two detectors provide different types of measurements.
The SVD delivers precise three-dimensional hits reconstructed from correlated clusters on opposite sides of the silicon strip sensors~\cite{adamczyk2022design} arranged in 4 layers.
In contrast, the CDC records drift times of the sense wires, from which only two-dimensional information in the plane perpendicular to the wire can be obtained.
These sense wires are arranged in 32 axial layers aligned with the solenoidal magnetic field or in 24 stereo layers, with an angle of 45.4 to 74\, mrad in positive and negative direction in respect to the axial layers.
The longitudinal hit position in the CDC is obtained by combining measurements from axial and stereo layers; however, resolving the three-dimensional hit position requires associating hits to a track candidate and applying a trajectory hypothesis that is consistent across layers~\cite{taniguchi2017central,dong2019calibration}.
\\ 
As a consequence of this heterogeneous detector layout, Belle~II employs separate track-finding algorithms for the SVD and the CDC~\cite{BelleIITrackingGroup:2020hpx}, that are shown in \cref{fig:baseline_tracking}.  
First, hits belonging to the same particle are grouped and an initial estimate of the track parameters is obtained.
These collection is called track seeds. 
Second, a track-fitting algorithm refines these track seeds, yielding the final (HLT) tracks.
Track seeds are first formed in the CDC from the CDC hits with a CDC-only track finding algorithm.
Afterwards, these tracks are extrapolated inward to the SVD with a Combinatorial Kalman Filter (CKF)~\cite{ckf_ref}, and matched to the SVD hits. 
A second SVD-only track finding step is then performed on the remaining SVD hits, with the resulting new track seeds subsequently being extrapolated back to the CDC to attach remaining hits. 
This sequential, multi-stage track finding across these detectors is difficult to optimise and prone to mismatches, ultimately degrading track purity.
\\ \newline
Graph neural networks~(GNNs), in comparison to traditional algorithms, offer a solution for combining inputs from detectors with irregular geometries~\cite{Shlomi:2020gdn,Wang:2018nkf, Qasim:2019otl}. 
Our previous work, the CDC-only \textit{CDC AI Track (CAT) Finder}~\cite{cat_paper} showed strong performance for CDC track finding.
Since it replaces only the CDC-specific stage of the \belle reconstruction chain (\cref{fig:baseline_tracking}), the reconstruction remains staged and continues to suffer from SVD-CDC matching inefficiencies.
In this work, we extend our track finding algorithm and introduce the \textit{Belle II AI (BAT) Finder}, a unified multi-modal track finder that incorporates SVD and CDC hits within a single GNN architecture.  
The \bat extends the approach to a heterogeneous detector environment and reconstructs tracks in a single inference step. 
This paper describes the multi-modal model track finding including the architecture, dataset and implementation in \cref{sec:multimodal}, followed by the performance evaluation in \cref{sec:performance}, concluding in \cref{sec:conclusion}.

\section{Multi-modal track finding}
\label{sec:multimodal}
Following the approach of \belle tracking, the track finder must provide an initial estimate of the track parameters (position, momentum, charge), an ordered list of associated hits, and an initial covariance matrix of the track parameters for the final fit.
\\
As discussed in \cref{sec:introduction}, the staged reconstruction used at Belle~II requires separate track finding in the CDC and SVD, followed by inward and outward extrapolations. 

\subsection{Graph neural network}
\label{sec:gnn}
\begin{figure}
    \centering
    \includegraphics[width=0.9\linewidth]{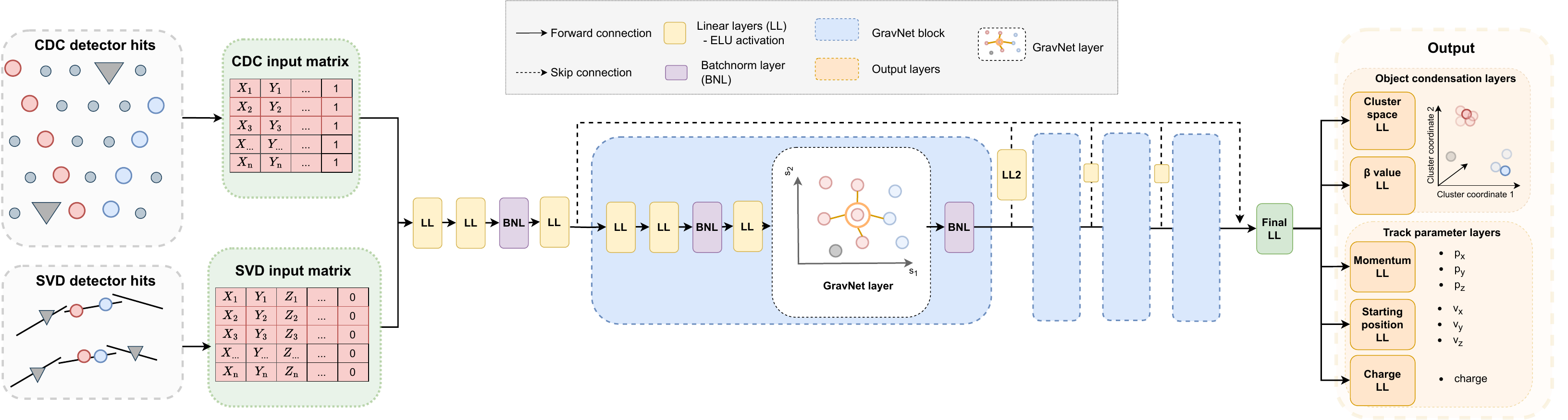}
    \caption{\bat model architecture.}
    \label{fig:model}
\end{figure}
The \bat model architecture is shown in \cref{fig:model}. 
It is adapted from the \cat~\cite{cat_paper} to incorporate inputs from both tracking detectors and is implemented in \pyg~\cite{Fey/Lenssen/2019}.
\\
All SVD and CDC hits are represented as nodes in a single graph, without any need for ordering or sequential approaches. 
The corresponding SVD input features include the global coordinates, deposited charge, timing, and signal-to-noise ratios of the two clusters.
CDC hit features are drift time, time-over-threshold, signal amplitude, and wire-geometry information required for three-dimensional reconstruction.
\\
The number of hits differs strongly between the two detectors. 
On average, tracks contain about four SVD hits for both high- and low-momentum particles, while CDC tracks contain roughly 50 hits for high-momentum particles and more than 200 hits for low-momentum particles. 
To account for this imbalance, the loss contribution of SVD hits is upweighted by an optimized factor of four during training.
\\
Hits from both detectors are provided to the network as a combined input matrix. 
Each hit is encoded by a fixed-length feature vector containing the union of all SVD- and CDC-specific quantities. 
Features not defined for a given detector are zero-padded, and a binary flag is used to encode the detector origin to either SVD (1) or CDC (0). 
\\
The feature vectors are mapped into a common latent space using fully connected layers and processed by GravNet blocks~\cite{Qasim:2019otl}.
Edges are constructed dynamically during message passing using the GravNet neighbourhood selection. 
No explicit subdetector or layer ordering is imposed. 
Instead, the network learns the relations between hits based on their features and local neighbourhoods, extending the approach previously used to combine the different geometries between axial and stereo CDC hits to also include SVD hits.
\\
The \bat utilizes the object condensation~\cite{Kieseler:2020wcq} loss function during training.
For each hit, the network predicts:
\begin{itemize}
    \item cluster-space coordinates used to group hits,
    \item a condensation score $\beta$ to identify track candidates,
    \item track parameters including initial momentum, start position, and charge.
\end{itemize}
During training, the loss function weights the parameter regression by the condensation score such that hits with high $\beta$ values are constrained to provide accurate track-parameter predictions, while predictions from low-$\beta$ hits are suppressed.
The loss function combines attractive and repulsive terms separate hits from different particles and builds track clusters from hits from the same particles in the cluster space.
In addition, a noise-suppression term is included to suppress background hits.
This approach removes the need for explicit SVD-CDC association logic, as hits from both detectors are grouped directly in the learned cluster space.

\subsection{Training strategy and evaluation datasets}
\label{sec:training}
The training strategy follows the approach established in the \cat~\cite{cat_paper} and is designed to optimize reconstruction performance across a wide range of track topologies.
The model is trained on a mixture of events that contain prompt and displaced tracks, with uniform sampling in charge, momentum, and vertex displacement to ensure a large variety of different final states, simulated using the \belle Analysis Software Framework~\cite{basf21, basf22}. 
Full detector background simulations are included, incorporating correlated noise across detector components to reflect realistic \belle operating conditions~\cite{Liptak:2021tog, Natochii:2022vcs}.
\\
Performance is evaluated using simulated events containing two displaced muon particles per vertex, where we sample between one and five vertices per event, defined in detail in~\cite{cat_paper}.
The two particles per vertex have opposite charges and momenta between 0.05\,GeV/$c$ - 6\,GeV/$c$ sampled uniformly over the full tracking acceptance.
Track starting points have uniformly sampled displacements \rhogenthreed up to 100\,cm, calculated in 3D  $\rhogenthreed = \sqrt{v_x^2 + v_y^2 + v_z^2}$.
For each particle, the momentum direction is sampled uniformly by rotating the radial vector from the origin to the starting point by an angle \opengen between 0$^\circ$ and 90$^\circ$ around a randomly chosen perpendicular axis.
For the performance evaluation in \cref{sec:performance}, the transverse displacement
\begin{equation}
    \rhogen = \sqrt{v_x^2 + v_y^2}
\end{equation}
is used, as it directly impacts the number of hits recorded in the tracking detectors.

\subsection{Implementation}
\label{sec:implementation}
During inference, hits with high condensation scores are selected as track seeds using score thresholds and local isolation criteria. 
For each seed, the network predicts the initial momentum, vertex position, and charge. 
Hits are then assigned to the nearest seed in cluster space, forming track candidates that include ordered hit lists and initial parameter estimates.
\\
These track candidates, consisting of CDC and SVD hits, are passed to the standard Belle~II track fitting with the GENFIT2 Kalman filter~\cite{Hoppner:2009af,Rauch:2014wta,Bilka:2019ang, GENFIT_zenodo}. 
For this study, energy loss is evaluated under the pion mass hypothesis for the track fitting, and a Deterministic Annealing Filter is used to down-weight hits with considerable discrepancy from the fitted trajectory~\cite{BelleIITrackingGroup:2020hpx}.

\section{Performance Results on Displaced Tracks}
\label{sec:performance}
\begin{table}[b]
    \centering
    \caption{Average track efficiency and track purity for displaced muons over the full detector region and their statistical uncertainties. 
    The uncertainties of the track finding algorithms are correlated, since they use the same simulated events.}
    \label{tab:performance}
    \begin{tabular}{lcc}
        \hline
        \textbf{Algorithm} & \textbf{Track efficiency} & \textbf{Track purity} \\
        \hline
        Baseline & $0.4804 \pm 0.0010$ & $0.9218 \pm 0.0009$ \\
        \cat & $0.6848 \pm 0.0008$ & $0.9380 \pm 0.0005$ \\
        \bat (this work) & $0.7469 \pm 0.0007$ & $0.9763 \pm 0.0004$ \\
        \hline
    \end{tabular}
\end{table}

\begin{figure}
    \centering
    \includegraphics[width=0.55\linewidth]{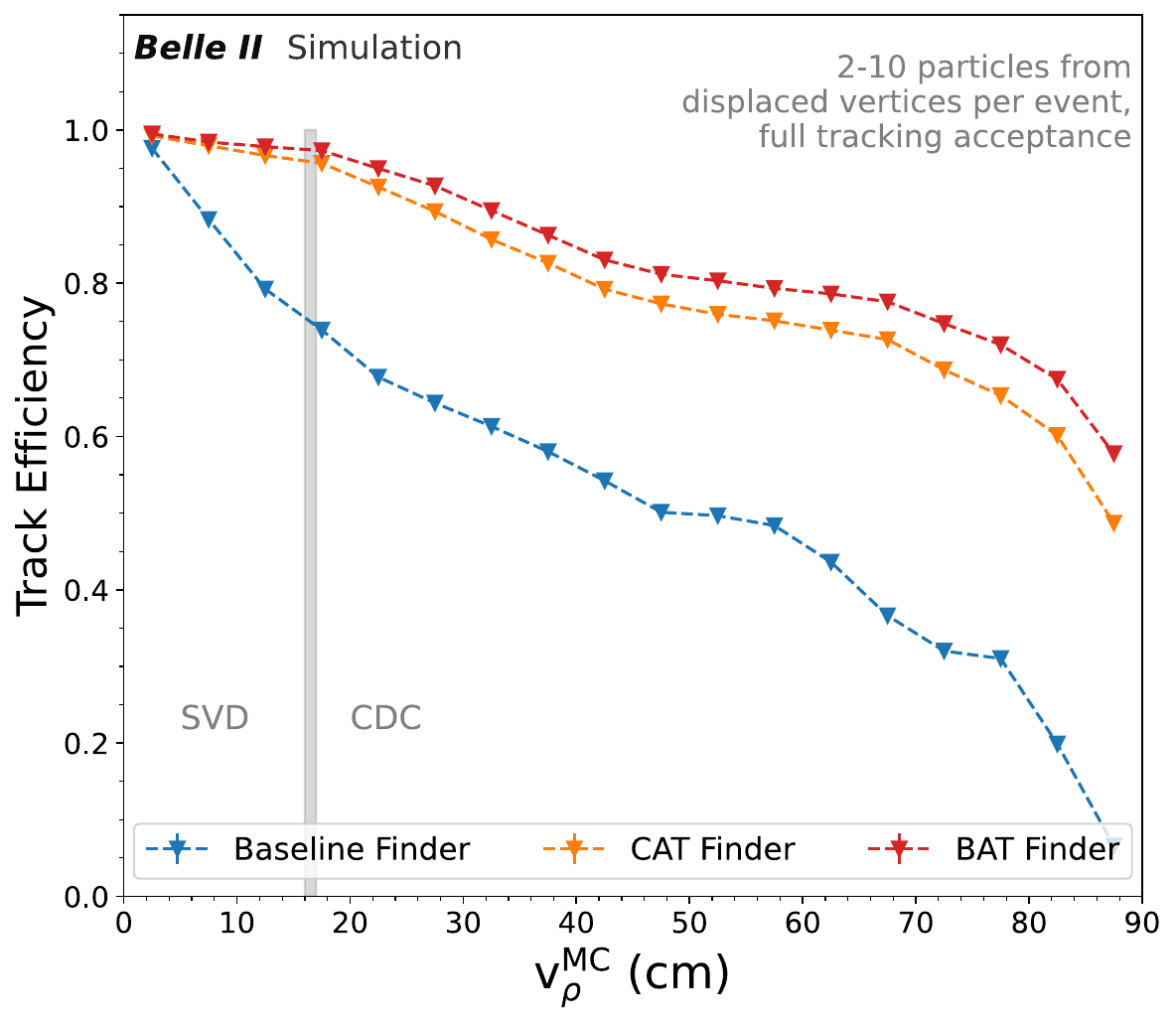}
    \caption{Track efficiency as a function of transverse displacement $v_{\rho}^{\mathrm{MC}}$ for the \legendre in blue, \cat in orange and \bat in red. 
    The grey vertical line marks the transition between the SVD and the CDC, corresponding to the outer radius of the SVD.
    For displacements beyond this radius, tracks originate inside the CDC and no longer traverse the SVD.}
    \label{fig:displaced_eff}
\end{figure}

The performance evaluation focuses on displaced tracks, a topology for which the \legendre shows continuously reducing efficiency as displacement increases, as shown in ~\cref{fig:displaced_eff}.
Track efficiency is defined as the fraction of unique tracks matched to particles over the number of charged particles that produce at least one hit in the SVD or CDC, while track purity is defined as the fraction of reconstructed tracks that can be matched to a true charged particle over all reconstructed tracks~\cite{BelleIITrackingGroup:2020hpx,cat_paper}.
\\
The averaged track efficiency and track purity for the \legendre, \cat, and \bat are summarised in Table~\ref{tab:performance}. 
Compared to both reference methods, the \bat achieves the highest track efficiency with 74.7\% compared to the 68.5\% of the \cat and 48.0\% of the \legendre. 
Most notably, the average track purity exceeds 97.6\% for the \bat, improving 5.5 and 3.8 percentage points over the \legendre and \cat, respectively.
This improvement stems from the single inference reconstruction approach, which replaces the staged SVD and CDC matching of the \legendre and \cat. 
By reconstructing tracks directly from the combined input of both subdetectors, the \bat minimizes failed hit associations and double reconstruction, which would otherwise generate duplicate tracks by the CDC and SVD track finding systems, leading to much cleaner track collections.
\\
Figure~\ref{fig:displaced_eff} shows the track efficiency as a function of transverse displacement.
While the \cat improves reconstruction for displaced tracks at larger radii, it slightly underperforms relative to the \bat for tracks with transverse displacements below 15\,cm, where it cannot fully leverage information from the region between the interaction point and the inner wall of the CDC. 
This is expected, as the \cat replaces only the CDC track finding stage and continues to rely on the baseline SVD reconstruction and staged SVD-CDC matching (\cref{sec:introduction}).
The \bat overcomes this limitation by incorporating SVD hits directly into the track finding, resulting in better performance for small and intermediate displacements.
Furthermore the \bat benefits from the inclusion of CDC time-over-threshold information, which was not utilized by the \cat~\cite{cat_paper} and improves hit discrimination between signal and beam-background.
Together, these effects result in a consistent efficiency gain of the \bat over the \cat and \legendre across the full displacement range shown in \cref{fig:displaced_eff}.

\noindent
Furthermore, the \bat maintains efficiency comparable to existing methods for prompt tracks while significantly enhancing track purity.
The high track purity demonstrates that the unified multi-modal reconstruction effectively suppresses fake tracks by reducing incorrect hit associations and avoiding the need for detector-to-detector track matching.

\section{Conclusion}
\label{sec:conclusion}
We presented the \bat, a unified graph neural network for track reconstruction at Belle~II that integrates inputs from both, silicon strips and a drift chamber, in a single inference step. 
By avoiding the staged CDC to SVD to CDC transition of the \legendre, which requires detector-to-detector matching that is difficult to optimize, our algorithm achieves high track efficiency while substantially improving track purity, reaching 97.6\% compared to the 92.2\% of the \legendre for muons from uniformly displaced vertices over the full tracking acceptance. 
\\
The results demonstrate that a single inference, multi-modal GNN approach can effectively operate on heterogeneous detector inputs and suppress misreconstruction that otherwise occurs during the subdetector matching for the \legendre and \cat. 
The \bat provides a robust solution for high-purity, high-efficiency tracking and offers a promising path for future high-luminosity \belle operations.

\bibliographystyle{unsrt} 
\bibliography{bibliography}

\end{document}